\documentclass[11pt]{article}
\usepackage{amsmath,amssymb,amsthm}
\usepackage{listings}
\usepackage{comment}
\usepackage{titlesec}
\usepackage[table,xcdraw]{xcolor}

\usepackage{graphicx}
\usepackage{subfigure}
\graphicspath{ {plots/} }
\usepackage{pdfpages}
\usepackage{csvsimple}
\usepackage{float}
\usepackage{caption}
\usepackage{url}
\usepackage{xcolor}[dvipsnames]
\newcommand{\bch}{\color{black}\em  }   
\newcommand{\ech}{\color{black}\rm  }    

\newcommand{\rd}{\color{black}}
\newcommand{\bk}{\color{black}}
\newcommand{\bl}{\color{black}}

\DeclareMathOperator*{\argmin}{\arg\!\min}
\DeclareMathOperator*{\argmax}{\arg\!\max}

\newcommand{\eqd}{\overset{d}{=}}

\renewcommand{\th}{\theta}
\newcommand{\lam}{\lambda}
\newcommand{\tht}{\tilde\theta}
\newcommand{\om}{\omega}
\newcommand{\omt}{\tilde\om}
\newcommand{\At}{\tilde{A}}
\newcommand{\Kp}{K^+}
\newcommand{\rb}{\bar{r}}

\newcommand{\Ktp}{\tilde{K}^+}
\newcommand{\Poi}{\mbox{Poi}}

\usepackage{natbib}

\usepackage{fullpage}

\title{A Recommender System Based on a Double Feature Allocation Model}
\author{Qiaohui Lin and Peter M\"uller}
\date{}

\begin{document}
\maketitle
\begin{abstract}
A collaborative filtering recommender system predicts
user preferences by discovering
 common features among users 
and items. We implement such inference using a
Bayesian double feature allocation model, that is, a model for random
pairs of subsets. We use an Indian buffet
process (IBP) to link users and items to features.
Here a feature is a subset of users and a matching subset of items.
By training feature-specific rating effects, we predict ratings.
We use MovieLens Data to demonstrate posterior inference
 in  the model
and prediction of user preferences for unseen items compared to
items they have previously rated.

Part of the implementation is a  novel 
semi-consensus Monte Carlo method to
accomodate large numbers of users and items, as is typical for related
applications. The proposed approach implements parallel posterior
sampling in multiple shards of users while sharing item-related global
parameters across shards.
\end{abstract}

\section{Introduction}
\label{sec:intro}
We develop a nonparametric Bayesian model-based approach to
collaborative filtering for random subsets of items and
users. The main contributions are the construction of a suitable prior
for pairs of subsets of items and users (features), the possibility to
report coherent inference on such features, and a consensus Monte
Carlo approach to allow practical implementation of posterior
inference.

Collaborative filtering refers to recommender systems that \ech predict
personalized user preferences for products (i.e, ratings, rankings) by
discovering similarity patterns among users and items, and make corresponding recommendations (\cite{sarwar2001item},
\cite{CFRS}, \cite{ACF}). It has been widely adopted by e-commerce
websites and online streaming services. The Netflix Prize since 2006
has encouraged more progress in this field.

The Netflix data (\cite{LDA-like}, \cite{ACF}, \cite{VCF1},
\cite{VCF2}) is a widely used benchmark
dataset used in collaborative
filtering research. It is a sparse matrix with 463435 rows for users,
17769 columns for items (movies) and 56.9 million
entries based on ratings between 1999 and 2005. On average a movie
has 5600 ratings and a user rates 208 movies (\cite{ACF}). Thus the
density of the matrix is as low as 0.69\% (\cite{VCF2}).

Collaborative filtering methods have progressed from naive nearest
neighbor methods, to well-adopted matrix factorization methods, to
probabilistic Bayesian models with latent factors and, more
recently, to generative models.
Nearest neighbour methods (\cite{CFRS}, \cite{ACF})  are intuitively appealing. The
idea is that the rating $r_{ui}$ of user $u$ for item $i$ is likely to be close to
the ratings $r_{u'i}$ of similar users $u'$ for the same item, or
the ratings $r_{ui'}$ of similar items $i'$ by the same user $u$. 
The key here is to measure the similarity  of  users (or
items). The simplest measure uses the correlation coefficient.
Using a similarity measure, we can identify the $k$ items that are
most similar to $i$ rated by $u$.
Denote the set of such items by $S^k(i;u)$. We can then predict $r_{ui}$ using a weighted average of the ratings of items in
$S^k(i;u)$ that are rated by user $u$, using weights proportional to
the respective similarlities, i.e, the more similar the neighbour is,
the more weight it gets.

Assume now that ratings are metric, facilitating the use of one of
the most popular methods in collaborative filtering based on
matrix factorization. The $m\times n$ rating matrix $R=[r_{ui}]$ with $m$ users
and $n$ movies can be written as the product of a $k\times m$
matrix $P$ and $k \times n$ matrix $Q$, i.e, $R=P^TQ$.
Here,  $k$ is the  number 
of latent factors $z$, for example, movie genres. Then
 the $P$ matrix can be interpreted as representing the preferences 
of users for the  $k$ genres and  the $Q$ matrix
can be interpreted as a classification of the movies relative to 
these genres.
The rating $r_{ui}$ for a specific movie and user then becomes
$r_{ui} = p_u^T q_i$, using $p_u$ and $q_i$ to denote the
corresponding columns of $P$ and $Q$, respectively. 

The problem then is to find preferences $P$ and $Q$ to best predict
observed ratings $R$ by $\hat{R} = P^TQ$, while controlling
the number of the latent factors (the rank of $P$ and $Q$). Thus the
objective is
\begin{equation}
  min_{ Q,P}\sum_{u,i}(r_{ui}-p_{ u}^Tq_{ i})^2+
   \lambda(\sum_i ||q_i||^2+ \sum_u ||p_{ u}||^2).
  \label{matFac}
\end{equation}
The  parameter $\lambda$ is determined by cross
validation. Minimization is performed by stochastic gradient
descent. A popular implementation is reported
in \cite{funk2006MF} and
\cite{Paterek2007MF}.

Matrix factorization captures the latent pattern in users and items. A
convenient stochastic gradient descent made it a winner of the Netflix
prize.  An additional advantage is the ease of incorporating temporal
dynamics.  However, the point estimation of the prediction comes
without an uncertainty measure and thus cannot serve a more
complicated goal of filtering and understanding user behavior, one of
the reasons why we introduce probabilistic models in the next section.

\section{Probabilistic Models}\label{sec:other-prob-models}
Probabilistic matrix factorization \citep{PMF} interprets
\eqref{matFac}
from a probabilistic perspective. They show that minimizing the sum of
square errors while penalizing their Frobenius norm is equivalent to
maximizing the log posterior in a probabilistic model with spherical
Gaussian priors.

 As before, define $R=[r_{ui}]$ as the $(m \times n)$ rating
matrix, and $P$ and $Q$ are $(k\times m)$ and $(k\times n)$ 
the low rank latent feature-user and feature-item matrices.
As before let $p_u$ and $q_i$ denote column $u$ and $i$ of $P$ and $Q$, respectively,
and let $N(x \mid m,V)$ denote a normal p.d.f for random variable $x$ with moments $m$ and $V$. 
We assume 
\begin{equation*} 
  p(R \mid P,Q,\sigma^2)=
  \prod^m_{ u=1}\prod^n_{i=1}[N(r_{ui} \mid p_u^Tq_i,\sigma^2)]^{I_{ui}}.
\end{equation*}  
with $I_{ ui}=1$ when user $u$ has rated movie $i$ and $I_{ui}=0$ otherwise.
The model is completed with zero-mean spherical Gaussian priors on $P$ and $Q$:
\begin{equation*}
    p(P \mid \sigma_P^2)=\prod^m_{u=1} N(p_u \mid
    0,\sigma_P^2I),\quad p(Q \mid \sigma_Q^2)=
    \prod_{i=1}^n N(q_i \mid 0,\sigma_Q^2I) 
\end{equation*}
Maximizing $\log p(P,Q \mid R,\sigma^2,\sigma_P^2,\sigma_Q^2)$
under this model is equivalent to minimizing the sum of squares error with quadratic
regularization, as in \bk
\begin{equation}\label{eq:mf-obj}
    E=\frac{1}{2}\sum_{u=1}^m\sum_{i=1}^nI_{ui}(r_{ui}-p_u^Tq_i)^2+\frac{\lambda_P}{2}\sum_{u=1}^m ||p_u||^2+\frac{\lambda_Q}{2}\sum_{i=1}^n ||q_i||^2,
\end{equation}
where $\lambda_P=\frac{\sigma^2}{\sigma_P^2}$, $\lambda_Q=\frac{\sigma^2}{\sigma_Q^2}$.

A minibatch gradient descent is used to find the optimal $P$ and
$Q$.  The optimization defines  an extension of the classic SVD
model, where the
 modified SVD is defined as the MAP estimate, 
and the classic SVD is a special case where prior variance goes to infinity.  

\cite{PMF} also discussed other constraints to be allowed onto the
model. For all proposed models, instead of adopting a full
Bayesian approach and leading to a MCMC posterior simulation, the
authors used stochastic gradient descent. This approach limits
meaningful posterior inference for hyperparameters and uncertainty
quantification of the MAP, but on the other hand has vastly decreased
the computational cost.    

Some probabilistic models proposed in the recent literature build on the Bayesian Mallows Model (\cite{liu19diverse}, \cite{liu19model}, \cite{vitelli17mallows}). Different from other
models, Mallows model  treats the response variables as ordinal
rankings. For example,  \cite{liu19diverse} work with the ranking on $n$ items for a user,
$R_u=\{R_{u1},R_{u2}, \ldots, R_{un}\}$, $R_{ui} \in \{1,\ldots,n\}$ 
and $u=1, \ldots, m$.
Mallows model is a  probability  model on the space $P_n$ of
permutations of $n$ items.
A basic model uses a latent consensus ranking $\rho \in P_n$ to
define,
\begin{equation}
  p(R_{u}=R \mid \alpha,\rho)  \propto
  \exp\left(-\frac{\alpha}{n}\, d(R,\rho) \right),
  \label{eq:mallows}
\end{equation}
where $\alpha$ is a scale parameter and
$d(R,\rho)=\sum_{i=1}^n|R_i-\rho_i|$ is a distance between $R$ and $\rho$.
 The normalizing constant 
$Z_n(\alpha,\rho)$ in \eqref{eq:mallows} 
is usually not analytically tractable. 
\cite{vitelli17mallows}, for example, use instead importance sampling and
Metropolis-Hastings posterior simulation schemes. 

In a more complicated scenario where users are not homogeneous, $m$ users are arranged into $C$ clusters; each cluster has its own common consensus $\rho_c$.  Latent cluster membership indicator $z_u$ assign user $u$ to cluster $z_u$. The model is then
\begin{equation}
  p(R_1,\ldots,R_m \mid z_1,\ldots,z_m,
  \alpha_c,\rho_c;\; c=1,\ldots,C)=
  \prod_{u=1}^m
  [Z_n(\alpha_{z_u})]^{-1}\exp\left\{-\frac{\alpha_{z_u}}{n}d(R_u,\rho_{z_u})\right\}.
  \label{eq:Mallows}
\end{equation}
with exponential priors on $\alpha_c$, uniform prior on $\rho_c$.  \rd The prior for the cluster assignments $z_u$, $u=1,...,m$ is $p(z_1,\ldots,z_u \mid \tau_1,\ldots,\tau_C)=\prod_{u=1}^m \tau_{z_u}$ with a Dirichlet prior on $\tau$. \bk

Assume now that a given user $u$ has rated only $m_u<m$ (instead of
all $m$) items and the objective is to make $L$ recommendations. This is equivalent to inferring \bch the unseen items with the $L$ highest rankings. Let $H_u$ denote the top $L$ rankings different from the rankings of the observed items. 
To find the top $L$ items for user $u$, we evaluate the posterior probability
\begin{equation*}
  p_{iu}=p(R_{ui} \in H_u \mid data)
\end{equation*} \ech
for all unrated items $i$
(skipping details of how model \eqref{eq:Mallows} is modified to
allow for the observation of $m_u<m$ items only). 
The strength of Mallows model is the use of a distribution of
rankings, allowing inference beyond point estimation. The main
limitation is the need for computation-expensive posterior MCMC
simulation, which is not suitable for large data sets on sequential
updating.

Some approaches to collaborive filtering are based on LDA (latent
Dirichlet allocation) type models. For example, the User Rating Profile model (URP) proposed in \cite{LDA-like}
represents each user as a mixture as a user attitudes, and the mixture
proportions are distributed according to a Dirichlet random variable.
For any user  $u$ we introduce a set of latent item-specific attitudes
$Z_{ui}$. Here $Z_{ui}$ is a user attitude that determines the rating
of $i$.
Next, let $\beta_{viz} = P(r_{ui}=v \mid Z_{ui}=z)$ denote
item-specific rating probabilities.
Like in the Latent Dirichlet
Allocation (LDA) model, $\theta_u$ is a Dirichlet random variable with
parameter $\alpha$, and $p(Z=z)=\th_{uz}$. 

A user profile is thus $r_u=[r_{u1}, \ldots, r_{un}]$
 with a sampling model
$p(r_u \mid Z_{u}, \beta)$ determined by the described hierarchical
model. 
However, the posterior distribution for $\theta$ and $Z_u$ is
intractable.
 Another line of research are based on the 
use neural networks or, more generally, generative models for
collaborative filtering
\citep{VCF1,cf-bandit,NCF,VCF2}.
Variational Autoencoder is one of the popular methods. 

The use of variational autoencoder models  originated from
\cite{VAE}. To apply it to the collaborative
filtering, following \cite{VCF2}, we assume the hidden factors are
latent variables $z_u$, the number of clicks of a user to all items is
$x_u=[x_{u1},,,x_{un}]$, $f_\theta(z_u)$ is a non-linear function to
produce a probability distribution of $\pi(z_u)$, and $x_u$ is
multinomial with total number of clicks $N_u$ and probability vector
$\pi(z_u)$.
\begin{align*}
    z_u \sim N(&0,I_k),\quad \pi(z_u)\propto \exp[f_{\theta}(z_u)],\\
    &x_{u} \sim \mathrm{Mult}(N_u, \pi(z_u)),
\end{align*}
One common approach is to use variational inference to approximate the
intractable posterior $p(z_u|x_u)$ with a variational approximate
distribution $q(z_u)$. Full algorithm can be seen in \cite{VCF2} and \cite{VAE}. 

Autorec model (\cite{VCF1}) is another version of this model applied to rating instead of clicking numbers. Generative Models with variational inference is an efficient scalable inference well suited when the dataset is large in size and new users' data keep flowing in. The problem is the lack of explanibity, the absence of uncertainty quantification, and the difficulty to fit the model to a more complicated goal, such as a specific criterion for recommendation diversity and accuracy tradeoff. 

Below we introduce a model that builds on these approaches, aiming to
(i) include learning on underlying structure that determines user
preferences, and (ii) still allows (approximate) full posterior
inference. The underlying structure that can be discovered by the
proposed model is an extension of the clusters that feature in Mallows
model by adding matching subsets of items.
%

\section{Double Feature Allocation Model}
\label{sec:model}
 We introduce an alternative generative model for user ratings
$r_{ui}$ of users $u=1,\ldots, m$ for items $i=1, \ldots, n$, using
notation as before. We assume ordinal ratings $r_{ui} \in
\{1,\ldots,5\}$. 
The inference   goal is to predict user preferences for  un-rated  movies by discovering
similarity patterns among users and movies.

The proposed model construction is guided by symmetry assumptions
with respect to items and users. That is, the
probability model should be invariant with respect to
arbitrary permutation of user indices, and/or permutations of movie
indices. Models with such structure are also known as separately
exchangeable, and the rating matrix can be characterized as 
\begin{align}\label{eq:SepExc}
    r_{1:m,1:n} \eqd r_{\pi_1(1:m),\pi_2(1:n)}
\end{align}
for separate permutations $\pi_1$ and $\pi_2$ of rows and columns, respectively.
\bch Here $X \eqd Y$ indicates equality in distribution for two random variables. \ech

\begin{figure}[hbt]
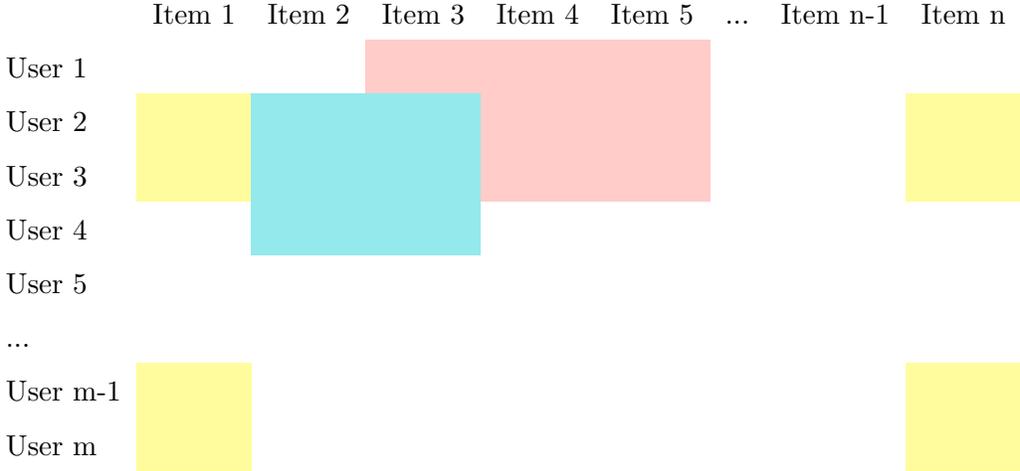

\begin{tabular}{lllllllll}
         & Item 1                   & Item 2                                          & Item 3                                          & Item 4                                          & Item 5                   & ... & Item n-1 & Item n                   \\
User 1   &                          & {\color[HTML]{333333} }                         & \cellcolor[HTML]{FFCCC9}{\color[HTML]{333333} } & \cellcolor[HTML]{FFCCC9}{\color[HTML]{333333} } & \cellcolor[HTML]{FFCCC9} &     &          &                          \\
User 2   & \cellcolor[HTML]{FFFC9E} & \cellcolor[HTML]{94E9ED}{\color[HTML]{333333} } & \cellcolor[HTML]{94E9ED}{\color[HTML]{333333} } & \cellcolor[HTML]{FFCCC9}{\color[HTML]{333333} } & \cellcolor[HTML]{FFCCC9} &     &          & \cellcolor[HTML]{FFFC9E} \\
User 3   & \cellcolor[HTML]{FFFC9E} & \cellcolor[HTML]{94E9ED}{\color[HTML]{333333} } & \cellcolor[HTML]{94E9ED}{\color[HTML]{333333} } & \cellcolor[HTML]{FFCCC9}{\color[HTML]{333333} } & \cellcolor[HTML]{FFCCC9} &     &          & \cellcolor[HTML]{FFFC9E} \\
User 4   &                          & \cellcolor[HTML]{94E9ED}                        & \cellcolor[HTML]{94E9ED}                        &                                                 &                          &     &          &                          \\
User 5   &                          &                                                 &                                                 &                                                 &                          &     &          &                          \\
...      &                          &                                                 &                                                 &                                                 &                          &     &          &                          \\
User m-1 & \cellcolor[HTML]{FFFC9E} &                                                 &                                                 &                                                 &                          &     &          & \cellcolor[HTML]{FFFC9E} \\
User m   & \cellcolor[HTML]{FFFC9E} &                                                 &                                                 &                                                 &                          &     &          & \cellcolor[HTML]{FFFC9E}
\end{tabular}
\vspace{2mm}
\caption{Illustration of features in Double Feature Allocation Model, with colored box as features.}
\label{fig:illustraion-of-DFA}
\end{figure}

A double feature allocation model is a model for random pairs of
subsets, first proposed in \citet{ni2019dfa}. \rd Our model is similar but different from the  Coupled Indian Buffet Process Model proposed by \cite{chatzis2012coupled}. \cite{chatzis2012coupled} used two independent Indian Buffet process (IBP) for two separate feature allocations of users and items. While in our case,  \bk a feature is a subset of
users  together with  a matching subset of items.  We use
an Indian buffet process 
(IBP)  prior to link  users and items to features. Suppose
there are $K$ features.
We use an  $(m \times K)$ binary matrix $A$ to link users to
features, with $A_{uk}=1$ indicating that user $u$ is in feature
$k$. 
Another $(n \times K)$ binary matrix $B$ links items to features,
 with $B_{ik}=1$ meaning that movie $i$ is in feature $k$. 
Figure~\ref{fig:illustraion-of-DFA} is a stylized representation of the double
feature allocation model. Same colored block indicate one feature
$k$, a subset of users and a matching subset of items.

The IBP prior on $A$ \citep{griffiths2011indian} is defined as follows. 
The model includes an unknown number $K$ of features, and can be written as
\begin{equation*}
    p(A)=\frac{\lambda^K\text{exp}(-\lambda H)}{K!}\prod_{k=1}^K\frac{\Gamma(m_k)\Gamma(m-m_k+1)}{\Gamma(m+1)},
\end{equation*}
where \rd $p(A)$ is without order of columns, \bk $\lambda$ is a fixed hyperparameter (concentration parameter),
$H$ is the harmonic number $H=\sum_{u=1}^m 1/u$, $m_k$ is the sum of
column $k$, $m_{k}=\sum_{u=1}^m A_{uk}$. The number of features $K$ is
random and unbounded, and features are exchangeable \bch a priori. \ech. The IBP is easiest described as a generative model building up $A$ row by row, starting with $u=1$, and adding columns of
$A$, i.e., features, \bch and indexing features by appearance. \ech
Let $K_u$ denote the number of
features that are introduced after the first $u$ users, starting with
$K_0=0$, and let $m_{u,k}=\sum_{v=1}^{u}A_{vi}$ denote the cardinality
of feature $k$ among the first $u$ users, $k=1,\ldots,K_{u}$.
Considering the respective next user $u$ we then proceed as follows.
First we decide inclusion into one of the existing features,
$k=1,\ldots,K_{u-1}$, with probability $p(A_{uk}=1 \mid A_{1\cdots
u-1, 1\cdots K_{u-1}}) = m_{u-1,k}/u$ Then we add a Poisson random
number $\Kp_u \sim \Poi(\alpha/u)$ new features with $A_{uk}=1$,
$k=K_{u-1},\ldots,K_{u-1}+\Kp_u$ and increment $K_u=K_{u-1}+\Kp_u$.
Implicit in the construction is a constraint of all zeroes in the right upper
corner of $A$, i.e., $A_{vk}=0$, $k>K_u$ and $v<u$.
We remove the constraint \bch of indexing items by appearance \ech by using a final step of randomly permuting
the final $K=K_n$ columns. 
In practice, for the purpose of prior sampling, we only need the
conditional probability of $A_{uk}=1$, 
\begin{equation*}
    p(A_{uk}=1|A_{-u,k})= m_{-u,k}/m,
\end{equation*}
where $A_{-u,k}$ is the $k$th column of $A$ excluding $A_{uk}$ and
$m_{-u,k}$ is the sum of column $k$ of $A$ excluding $A_{uk}$. 

Given $A$, the item-feature matrix $B$ inherits $K$ features from
$A$. For simplicity, we assume independent Bernoullis:   
\begin{equation*}
    p(B_{ik}=1|A)= p,
\end{equation*}
with the prior parameter $p$ usually chosen to be a small number to
control the number of features a movie can be in. Such parsimony
ensures that features do not share too many common movies and preserve
their differences. In posterior sampling,  $p$ and the dimension of
$B$ will be updated each time after we update $A$.  

Note that the IBP $p(A)$ together with $p(B \mid A)$ define a joint
model $p(A,B)$, and therefore also imply a marginal $p(B)$.  The
current model does not imply a marginal IBP prior for $p(B)$. If a
more symmetric construction with an IBP marginal prior on $B$ were
desired, it could be easily achieved.
Let $p_{\text{IBP}}(B)$ denote an IBP prior on a random binary matrix
$B$, including the number of columns, $K_B$, and let $p_{\text{IBP}}(B
\mid K_B)$ denote the conditional distribution of $B$, conditional on
the number of columns equal to $K_B$ under the IBP. Then using $p(B
\mid A) = p(B \mid K_B=K_A)$ would by construction deliver $p(B)=\text{IBP}$,
marginally.

Finally, we complete the inference model with a sampling model for the
observed ratings. The model links features to ratings by introducing
probabilities for $r_{ui}$ conditional on the currently imputed
features of which user $u$ and items $i$ are members.

We consider a baseline $b_0=2.5$ (between the extremes 1 and 5).
Each feature $k$  of which user $u$ and movie $i$ are a member
adds  an adjustment $\theta_k$ to this baseline.
For example, feature $k$  might be 
a pair of subsets of comedy movies and comedy fans. Then $\theta_k$
would be a positive increment from baseline. Similarly, if
feature $k$ is a pair of subsets of comedy movies and comedy haters,
then $\theta_k$ should be a negative adjustment. We allow an item and a
user to be in multiple subsets with features having aggregative
influences.

We also include global parameters $\rho_i$ for each movie regardless
of feature allocation. The parameter $\rho_i$ has an interpretation as
overall mean rating for movie $i$ (on the ordinal probit scale).  This
movie-specific offset reflects if a movie is generally popular and
well-received among audiences or vice versa. We complete the prior model with independent priors for $\th_k$
and $\rho_i$, assuming $\th_k \sim h_\th$ and $\rho_i \sim h_\rho$,
i.i.d. 

The sampling model is then defined as an ordinal probit model including the described feature-specific and
item-specific parameters.
Denote by  $Z_{ui}$  a latent continuous probit score for the
rating of user $u$ for movie $i$, denote by $K^*_{ui}=\{k:\;
A_{uk}=B_{ki}=1\}$ the set of features that include both, user $u$ and
movie $i$. 
We use an inverse-gamma hyperprior on $\tau^2$, and normal priors for
both $\theta$ and $\rho$.
We assume 
\begin{equation*}
  Z_{ui} \mid
  A,B,\theta,\tau \sim N(b_0+\sum_{k \in K^*}\theta_k+\rho_i,\tau^2).
\end{equation*}
The probit scores are linked with the data in the usual ordinal probit
model as 
\begin{align}\label{eq:probit-to-rating}
    r_{ui}= \left\{\begin{array}{cc}
       1,  &  Z_{ui}\leq1\\
       x, x\in\{2,3,4\}, & x-1 < Z_{ui} \leq x\\
       5, &  Z_{ui}>4
    \end{array}\right..
\end{align}

\section{Posterior Inference}
\label{sec:posterior-sampling}
\subsection{Posterior Sampling Algorithm}
\label{subsec:posterior-algorithm}
We implement posterior inference using MCMC posterior simulation. Let
$\omega=(A,B,\theta,\tau,\rho)$ denote the currently computed
parameters. See the appendix for transition probabilities to update
$B$, $\theta$, $\rho$ and $\tau$. Only the transition probability for
$A$ requires more discussion. 

Denote the number of iteration as superscript $^{(t)}$, denote the $u$th
row of $A$ and $R$ as $A_{u\cdot}$ and $R_{u\cdot}$, respectively, and denote the $i$th column of $R$
as $R_{\cdot i}$. The following three steps define a reversible jump transition
probability for $A_{u\cdot}$. Below, let $(\At,\tht)$ denote proposed
new values for $A,\th$, and let $\omt=(\At,B,\tht,\tau,\rho)$.
The transition probability to update $A_{u\cdot}$ is defined as
follows. 
\begin{enumerate}
\item For all the $k$ with $m_{-u,k} \neq 0$,
  update $p(A_{uk}=x \mid \cdot)\propto \frac{m_{-u,k}}{m}
          p(R_{u\cdot} \mid A_{uk}=x, A_{-u,k}, \theta,\rho,\tau,B)$, $x=\{0,1\}$. 
\item Reversible jump proposal.
  We refer to all features with $m_{-u,k}=0$ as {\em singular
    features}.
  W.l.o.g. assume $k=1,\ldots,K_0$ are not singular, and
  $k=K_0+1,\ldots,K$ are the singular features.
  We create a proposal by first dropping all singular features, i.e., retaining in $\At$ only the first $K_0$ columns of $A$,
  proposing $\At=A[\; \cdot \;, (1, \ldots, K_0)]$.
  Next we propose
  $\Ktp_u \sim Pois(\lambda/n)$ new (singular) features, together
  with (new) feature-specific parameters $\tht_k \sim h_\th$, $k=K_0+1,\ldots,K_0+\Ktp_u$.
  We add the new features to $\At$ with
  $\At_{uk}=1$ and $\At_{vk}=0$ for $k=K_0+1,\ldots,K_0+\Ktp_u$ and
  $v\ne u$.
\item  Metropolis-Hastings acceptance probability. We denote with $S=\{K_0+1,\ldots,K\}$ the indices of the singular features, and find 
 \begin{equation*}
    \alpha =
    \frac{p(R_{u\cdot} \mid \omt)}{p(R_{u\cdot} \mid \om)} \cdot
    \frac{p(\tht_{S}) \Poi(\Kp_u \mid \frac{\lam}{n})}
         {p(\th_{S})\, \Poi(\Ktp_u \mid \frac{\lam}{n})} 
    \cdot
    \frac{p(\th_{S}) \Poi(\Ktp_u \mid \frac{\lam}{n})}
         {p(\tht_{S})\, \Poi(\Kp_u \mid \frac{\lam}{n})}
    =
    \frac{p(R_{u\cdot} \mid \omt)}
         {p(R_{u\cdot} \mid \om)}
         \nonumber
       \end{equation*}
\end{enumerate}       
With probability $\min(1,\alpha)$, we accept the proposal
$\At,\tht$ and set $A=\At$, $\th=\tht$.
Otherwise we keep $A,\th$ unchanged.
 
We evaluate fitted mean ratings as
$\rb_{ui} = \int p(r_{ui}^{mis}=x|\omega) p(\omega|R^{obs})d\omega
\approx \frac1T \sum_{t=1}^T p(r_{ui}^{mis}=x|\omega^t)$
and predict unseen ratings for items $i$ with $I_{ui}=0$ by maximizing 
\begin{equation}\label{eq:predict-rating}
  \hat{r}_{ui}=  \argmax_{x\in \{1,2,3,4,5\}} \rb_{ui}
\end{equation}

\subsection{A Consensus Monte Carlo Method for Large Number of Users}
\label{subsec:concensus-monte-carlo}

The described posterior simulation can be computationally costly. 
Large recommender systems usually have at least thousands, or millions of users, render the usual MCMC impractical. Thus we use the idea of Consensus Monte Carlo \citep{scott2016bayes-consensus},  to  split data into shards, run MCMC one each shard in parallel on different machines, and then  reconcile posterior inference from the shards into a reconstruction of posterior inference under the full data.  \citet{ni2019consensus} applied Consensus Monte Carlo to Bayesian nonparametric models on clustering and feature allocation. Their method relies on shared data points (anchor points) across shards, and they merge random subsets (clusters or features) on different machines based on the number of common anchor points in the two sets. This strategy, though appealing, does not work in our case. \newcommand{\rhot}{\tilde\rho}
\newcommand{\sig}{\sigma}

Using a set of users as common anchors to define a criterion for
merging features would ignore the possibility of different sets of
movies being paired with these users in different shards. 
And similarly for using sets of movies only. A practicable
implementation would need to use a criterion based on shared users,
shared movies and similar imputed feature-specific effects $\th_k$.
A related criterion to merge subsets would require several ad-hoc choices and tuning parameters. In simulations we found that the involved approximations left the joint posterior reconstruction of little practical value. Instead we propose an alternative strategy based on Consensus Monte Carlo for global parameters, but keeping inference for random subsets local to each shard. We refer to this strategy as "semi-local Consensus Monte Carlo".

We use shards that split the data by subsets of users. 
Let $s=1, \ldots, S$ index the shards, and let $\bigcup_{s=1}^S U_i =
\{1, \ldots, m\}$ denote the split of users into the $S$ shards.
Also, let $R_s$ denote the data for the users in shard $s$.
In our model, the random features that are imputed under posterior
inference under shard $s$ naturally include only users from $U_s$,
making the features local parameters, while the only global
parameters are $\rho_i, i=1, \ldots, n$.
We use a CMC approximation of the joint posterior for the global
parameter $\rho$ as
$$
   p(\rho \mid R) \approx \prod_{s=1}^s p(R_s \mid \rho)\,
   p(\rho)^{1/S}.
$$
The nature of the approximation is to assume independence of the
marginal distribution of global parameters across shards. 
Conditional on a posterior sample $\rhot \sim p(\rho \mid R)$ we then
\bch use shard-specific posterior samples of the shard-specific parameters. \ech The
latter is implemented by selecting stored posterior Monte Carlo samples $(A,B,\theta,\rho,\tau)$ with $|\rho-\tilde{\rho}|<\epsilon$.
Here $A$ refers to users in shard $s$ only, and $B$ and $\theta$ are
linked with the subsets that are represented by $A$, making
$(A,B,\theta)$ parameters that are local to each shard only.

There remains the step of creating a Monte Carlo sample for
$\rho \sim p(\rho \mid R)$. For this we use Consensus Monte Carlo for approximate normal
posterior distributions. Let $(\mu_s, \sigma_s)$ denote posterior mean
and standard deviation of $\rho_i$ in shard $s$ (for a movie $i$).
We approximate $p(\rho \mid R) \approx N(\mu, \sig)$ with
$1/\sig^2 = 1/\sig^2_0 + \sum_s 1/\sig^2_s$ and $\mu = (\mu_0/\sigma_0^2+\sum_s\mu_s/\sigma^2_s)/(1/\sig^2_0 + \sum_s 1/\sig^2_s)$,
where $(\mu_0, \sig^2_0)$ are the prior moments for $\rho$. We call this strategy semi-local Consensus Monte Carlo.

The full algorithm is now, run MCMC described in
Section~\ref{subsec:posterior-algorithm} on each separate shards,
store $A$, $B$, $\theta$, $\tau$, $\rho$, every 5 iterations. Merging
\bch the shard-specific posterior distributions for $\rho$ across shards as described, we get an approximate global posterior from which we then generate a posterior Monte Carlo draw \ech
$\tilde{\rho}$. In each shard, for each stored iteration $A$, $B$,
$\theta$, $\tau$, resample $\tilde{\rho}$ from the global
distributions, filter the iterations where the stored shard-posterior
$\rho$ is close to resampled global $\tilde{\rho}$ and make
predictions based on $A$, $B$, $\theta$, $\tau$ and $\tilde{\rho}$ in
these iterations. Note that the filtering step within the shard is to
ensure the closeness of shard posterior MCMC and global posterior MCMC
and to not totally lose the dependence of $\tilde{\rho}$ and
$(A,B,\theta,\tau)$ in the iterations we use for prediction.  

\vspace{5mm}

Consensus Monte Carlo Algorithm
\begin{enumerate}
    \item Separate users into S shards, keep entire list of movies. 
    \item In each shard, \bch carry out MCMC simulation \ech for $A$, $B$, $\mathbf{\theta}$, $\mathbf{\tau}$, $\mathbf{\rho}$ according to the transition probability in Section~\ref{subsec:posterior-algorithm} and appendix. Store after thinning.
    \item Merge \bch the shard-specific posterior distributions for $\rho$, and derive an approximate global posterior distribution for $\mathbf{\rho}$ by aggregating shard precision. 
    \item In each shard, for each stored iteration, resampling $\mathbf{\tilde{\rho}}$ from the (approximate) global posterior, keep those iterations for which $|\rho-\tilde{\rho}|<\epsilon$, perform prediction using  $A$, $B$, $\mathbf{\theta}$, $\mathbf{\tau}$ and resampled $\tilde{\rho}$ in those iterations. 
\end{enumerate}

\section{Simulation}\label{sec:sim}
We implement the proposed scheme in R. The code is avaiblable in author's github. We use this to set up the simulations. In the simulation study, we generate a $100 \times 150$ rating matrix of $m=100$ users and $n=150$ movies from our model. We first simulate $A$ \bch under the \ech IBP prior with hyperparameter $\lambda=3$. The (random) number of columns of $A$ determines the number of features $K$. Next we simulate a $n \times K$ binary matrix $B$ with $p$, \bch using independent Bernoulli draws with success probability $0.2$. \ech
For user $u$ and movie $i$, identify the subset of features $K_{ui}^*={k: A_{uk}=B_{ki}=1}$. With $b_0=2.5$, $\theta_{1:K} \sim N(0,2)$ and $\tau=0.25$,  the latent probit score for rating, $Z_{ui}$ is simulated from 
\begin{align*}
    Z_{ui}|A,B,\theta,\tau \sim N(b_0+\sum_{k \in K^*_{ui}} \theta_k,\tau^2).
\end{align*}
Note here we did not introduce global parameter $\rho_i$ for movies as the simulation data size is small and we are not using Consensus Monte Carlo to split users into shards. Thus a global parameter for movies is optional but not necessary in this example. \bk The rating is generated from the probit score as in Eq~\ref{eq:probit-to-rating}.

\begin{figure}
    \centering
    \includegraphics[width=0.3\textwidth]{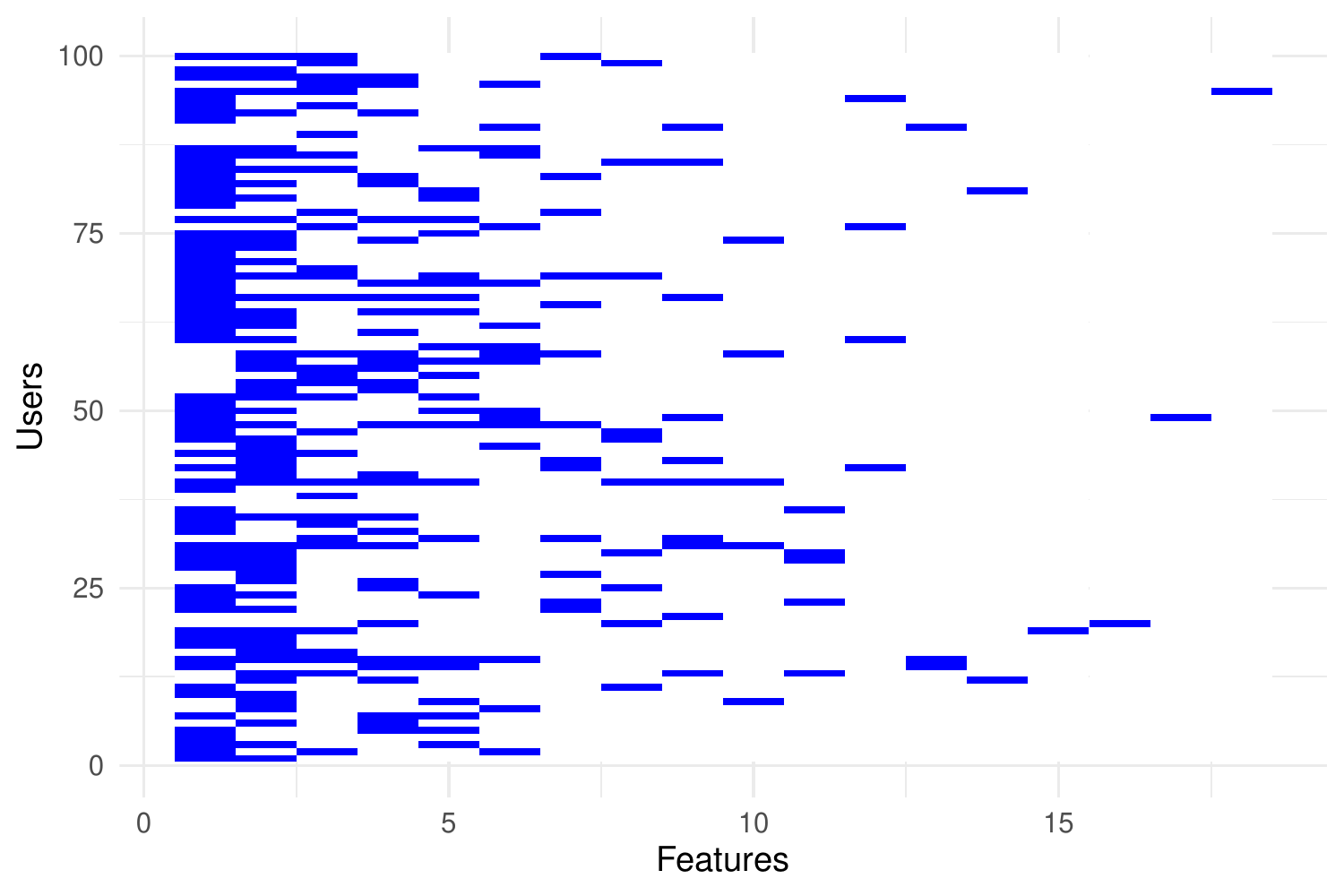}
    \includegraphics[width=0.3\textwidth]{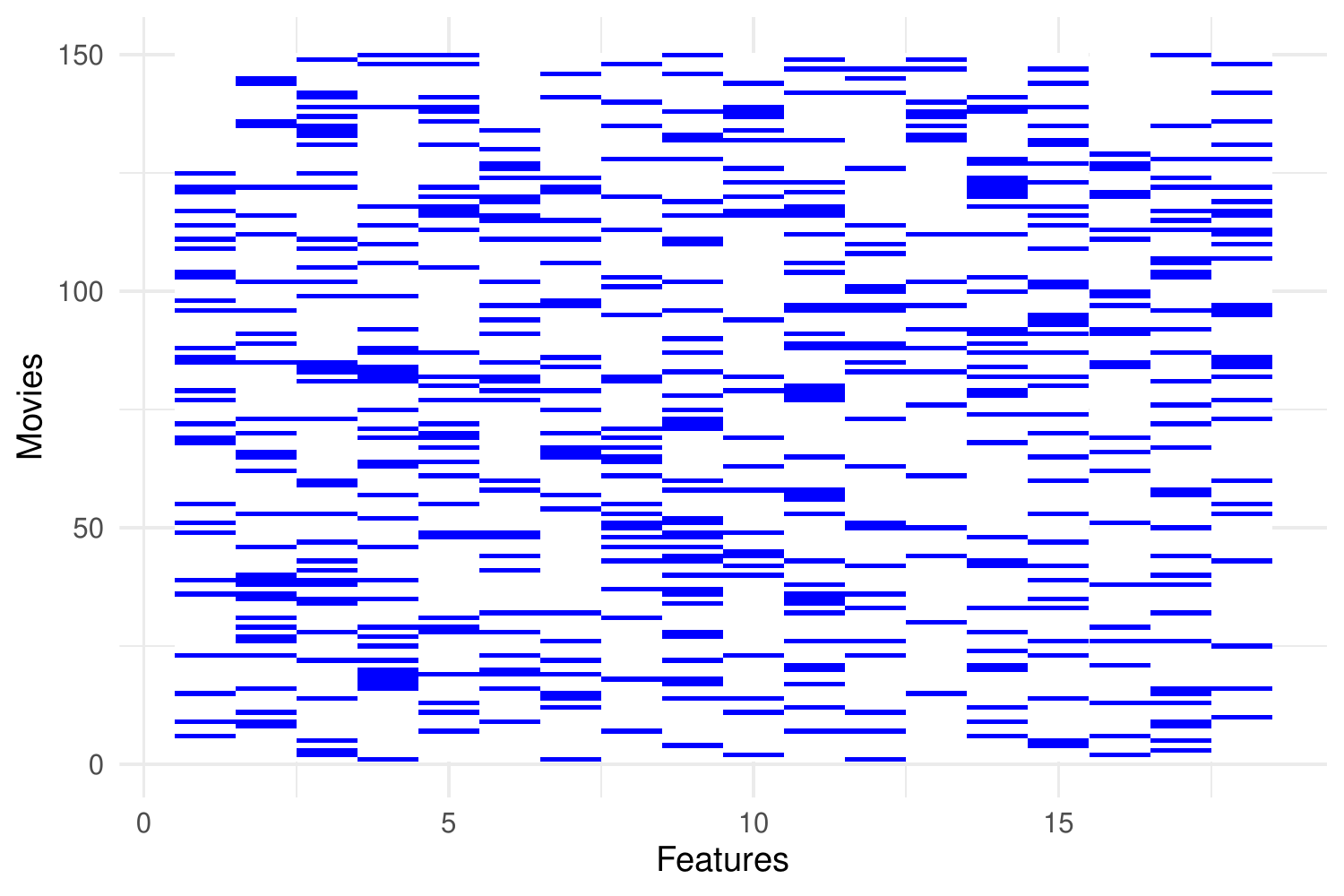}\\
    \includegraphics[width=0.3\textwidth]{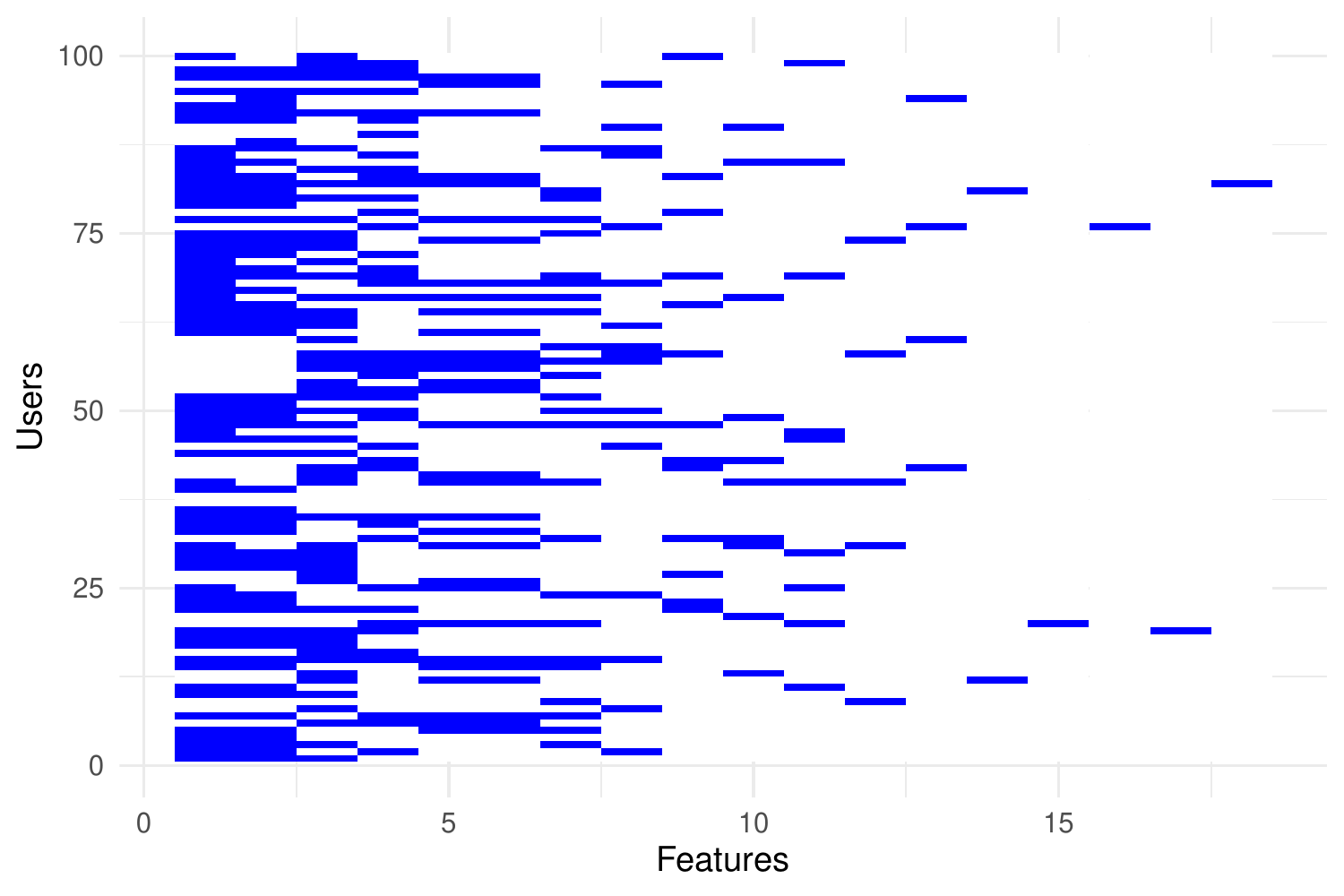}
    \includegraphics[width=0.3\textwidth]{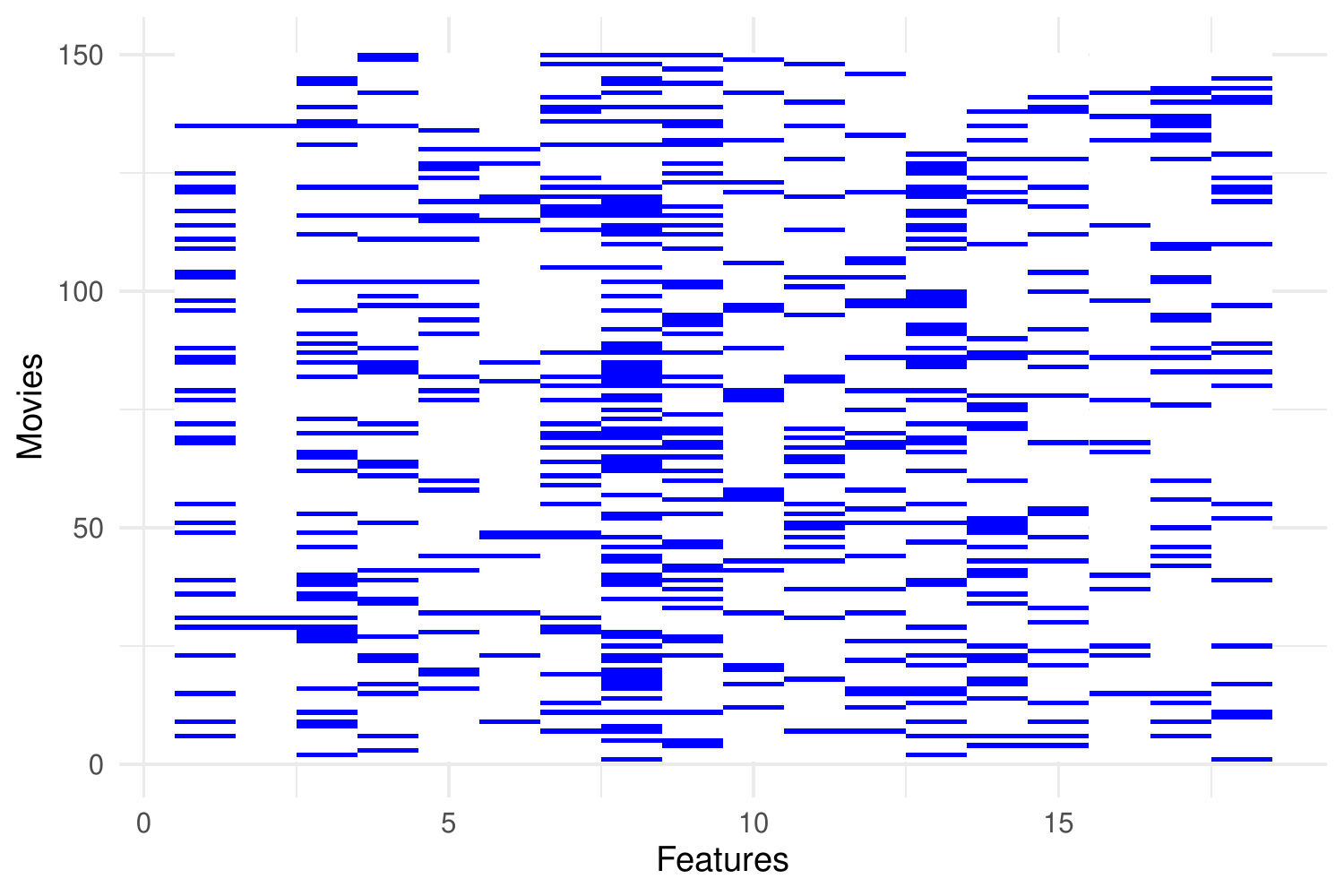}
    \caption{Simulation truth (top row) and posterior estimates (bottom row) for $A$, $B$. 
    Total feature (column) numbers are the same in the truth and estimates, but features (columns) can be permuted. }
    \label{fig:sim-ABtheta}
\end{figure}

We randomly split the simulated data into $80\%$ and $20\%$ for training and testing. We use the MCMC algorithm proposed in Section~\ref{subsec:posterior-algorithm}, implementing $10000$ iterations conditional on the training data. We evaluate an estimated user-feature relationships $\hat{A}$ following \cite{ni2019dfa}. We first calculate the maximum a posteriori (MAP) estimate $\hat{K}$ from the marginal posterior distribution of K. Conditional on $\hat{K}$, we follow \cite{dahl2006model} and compute a point estimate $\hat{A}$ as
\begin{equation*}
    \hat{A}=\argmin_{A'}\int d(A,A') dp(A|Z,R,\hat{K}),
\end{equation*}
where $d=\min_{\pi} H(A,\pi(A'))$, denotes the minimum Hamming distance between binary matrices $A$ and $A'$ over greedy searches of permutations of $\pi(A')$. In Figure~\ref{fig:sim-ABtheta}, we show the estimated user-feature relationships $\hat{A}$, and conditional on $\hat{A}$, the point estimates for movie-feature relationships $\hat{B}$ and the rating adjustment for each feature $\hat{\theta}$, versus the true values $A$, $B$, $\theta$ used in simulation. 

We predict rating for training and testing data by maximizing posterior predictive probability in Eq~\ref{eq:predict-rating}. The predicted rating for the training data set \bch is $69.12\%$ correct, meaning that in $69.12\%$ of the cases the predicted rating exactly matches the recorded data. The same for the test data was $66.89\%$. \ech.  We compare to inference under the matrix factorization method as described in Section~\ref{sec:other-prob-models}. We use the R package recosystem. Rank $k$ and sparsity parameters in Eq~\ref{eq:mf-obj} are tuned by cross validation. After training, we find a test RMSE of 0.67 and prediction accuracy for the test data set of $60.80\%$. 

\begin{figure}
\vspace{-5mm}
    \centering
    \includegraphics[width=0.45\textwidth]{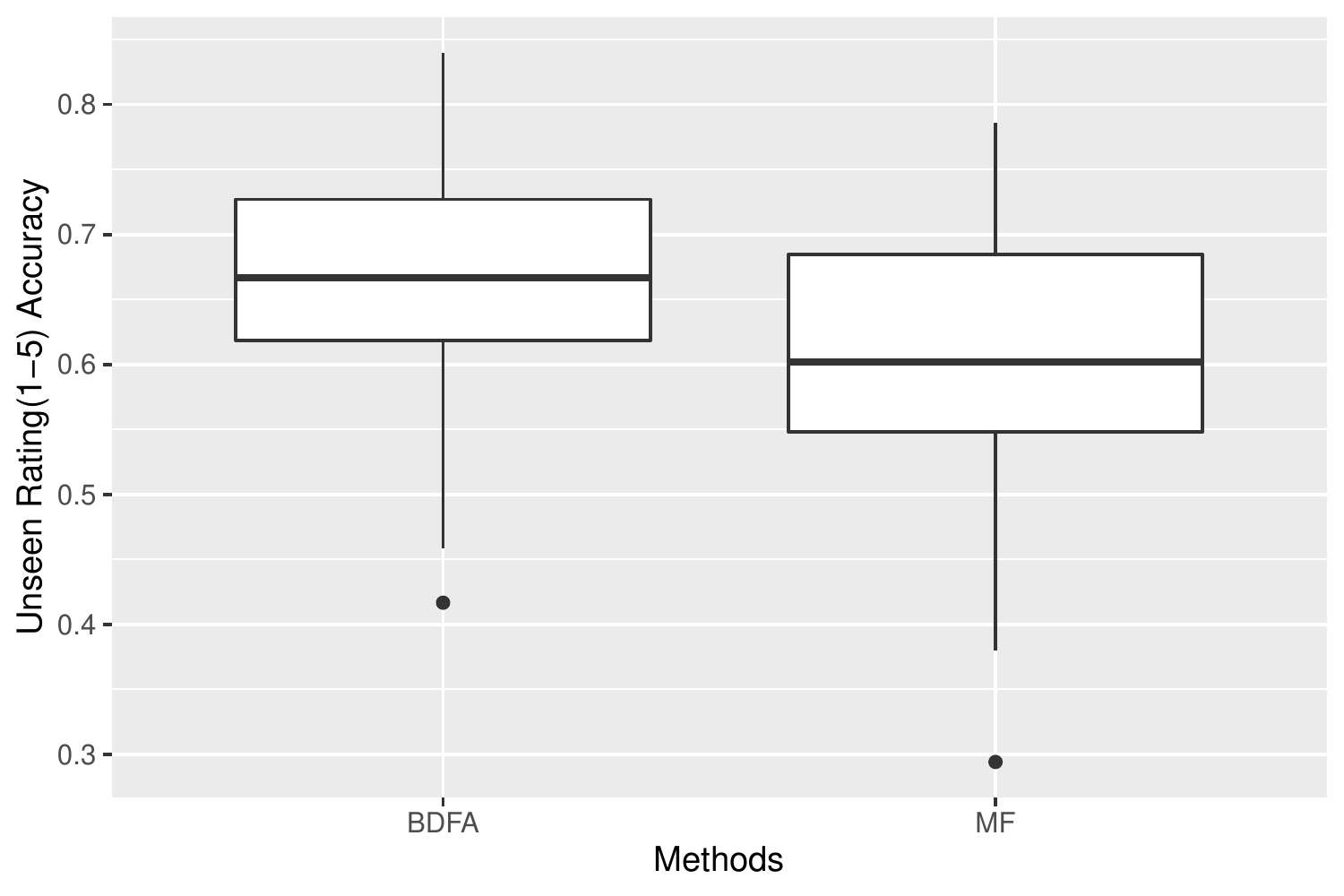}
    \includegraphics[width=0.45\textwidth]{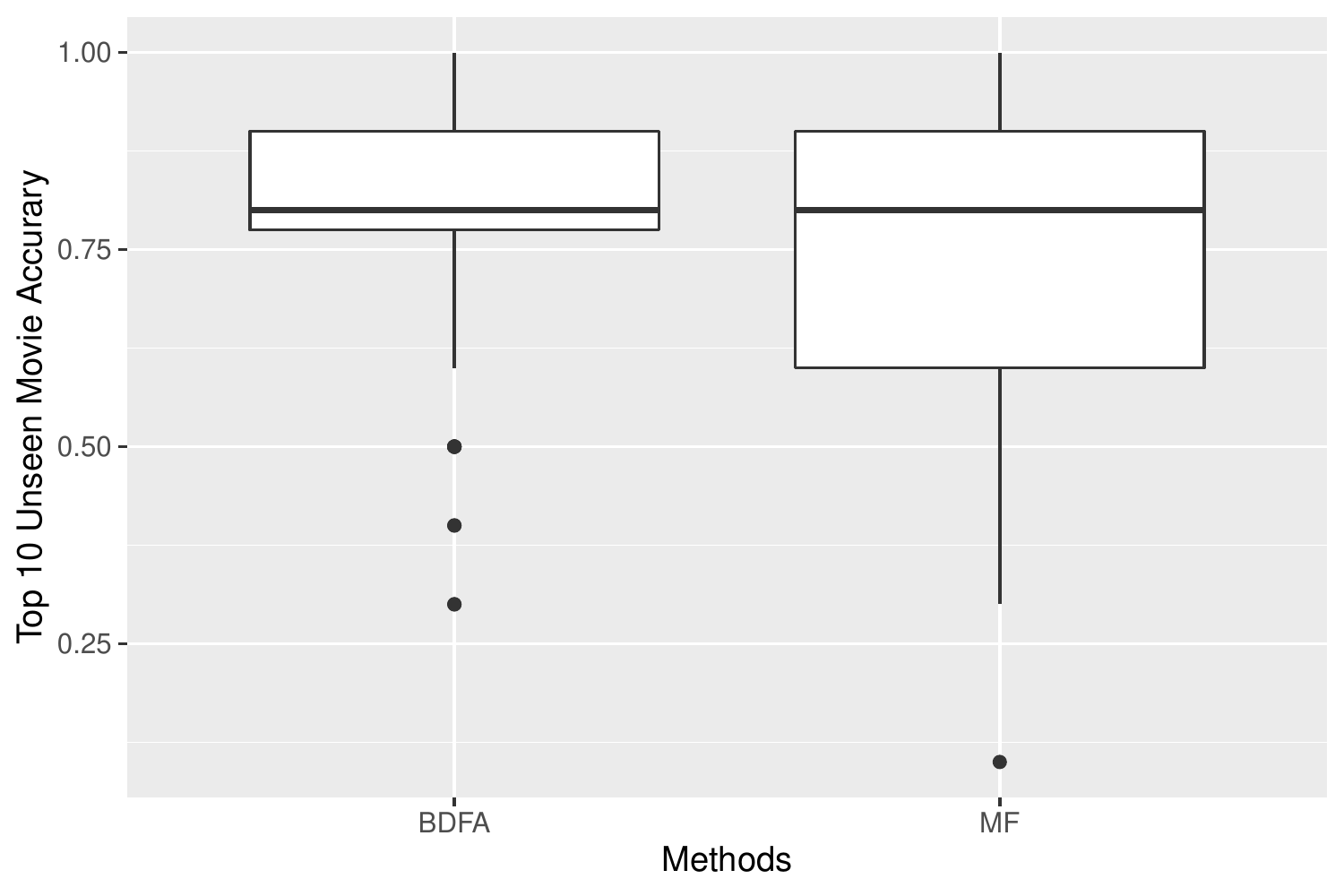}
    \vspace{-2mm}
    \caption{Boxplots of prediction accuracy for our method (Bayesian Double Feature Allocation, BDFA) to Matrix Factorization (MF). Left shows boxplot of all users' accuracy of predicting ratings (level 1-5) of unseen movies. Right shows boxplots of all users' accuracy of predicting the top 10 unseen movies.}
    \label{fig:sim-boxplots}
\end{figure}

Figure~\ref{fig:sim-boxplots} shows the boxplots of users' prediction accuracy for our methods and Matrix Factorization. The left boxplot shows all users' accuracy of predicting ratings (level 1-5) of unseen movies. The right boxplot shows all users' accuracy of predicting the top 10 unseen movies. Our method has a better simulation results in both boxplots.

\section{Preference Prediction with the MovieLens Data}
\label{sec:experiment}
We use the Movielens dataset \footnote{https://grouplens.org/datasets/} with movie ratings from 6040 users.  We clean the data \bch as in \ech  \citet{vitelli17mallows}, to keep results comparable.  We keep the $200$ most rated movies and users who rated more than three movies, which yields to a $6040\times 200$ rating matrix $R$. Our goal include traditional goals \bch for \ech recommender systems, such as predicting individual user's ratings to unseen movies, find the top-rated movies across all users, and a pairwise preference prediction goal similar to \cite{vitelli17mallows}. The latter goal predicts the preference of unseen movies to rated movies for each user. Here preference is defined as the following. 
Denote the predicted rating of user $u$ for movie $i$ as $\hat{R}_{ui}$, estimated as described in Section~\ref{sec:model}. We say that user $u$ strictly prefers movie $i$ to $i'$, and write $i \succ i'$ for user $u$, if $\hat{R}_{ui} > \hat{R}_{ui'}$. 

In the $6040\times 200$ rating matrix, only $24.7\%$ entries are observed. We conduct a pairwise preference prediction test using our model. For each user, we randomly select one movie that he/she has rated as test, and train \bch 
the model based on the remaining data for that user.  We compare the user's  preference of this test movie to other (rated) movies based on the predicted rating for the test movie to the observed ratings in the training set. \ech 
This pairwise preference test is also comparable to \citet{vitelli17mallows}.

We apply a double IBP prior described in Section~\ref{sec:model} initialized with $\lambda=3$, base line $\theta=2.5$, non-baseline $\theta$s drawn from $N(0,\sigma_0^2)$,  $\sigma_0=2$, $\tau$ drawn from an inverse-gamma distribution with location and scale parameters $(5,1)$.  We initialize the number of features $K$ from matrix factorization algorithm result. 

\bch Using  \ech the concensus Monte Carlo algorithm, we divided the $6040$ users into $15$ shards and \bch implement MCMC posterior simulation for each shard separately \ech in parallel. Each shard with $400$ users and $200$ movies stabilizes at around $39$ to $43$ features. 
As described in Section~\ref{sec:posterior-sampling}, we merge \bch inference across shards by defining the approximate global posterior for $\rho$.
For simplicity, we set the prior precision for $\rho$ to $0$, i.e., $\sigma_0=\infty$ and simplify the merge step to evaluating $\mu=\bar\mu_s$ as an unweighted average across shards. \ech The after-merge prediction follows from re-sampling $\rho$ from global posterior and previously stored MCMC draws of other variables. 

\begin{figure}
\vspace{-5mm}
    \centering
    \includegraphics[width=0.45\textwidth,height=0.35\textwidth]{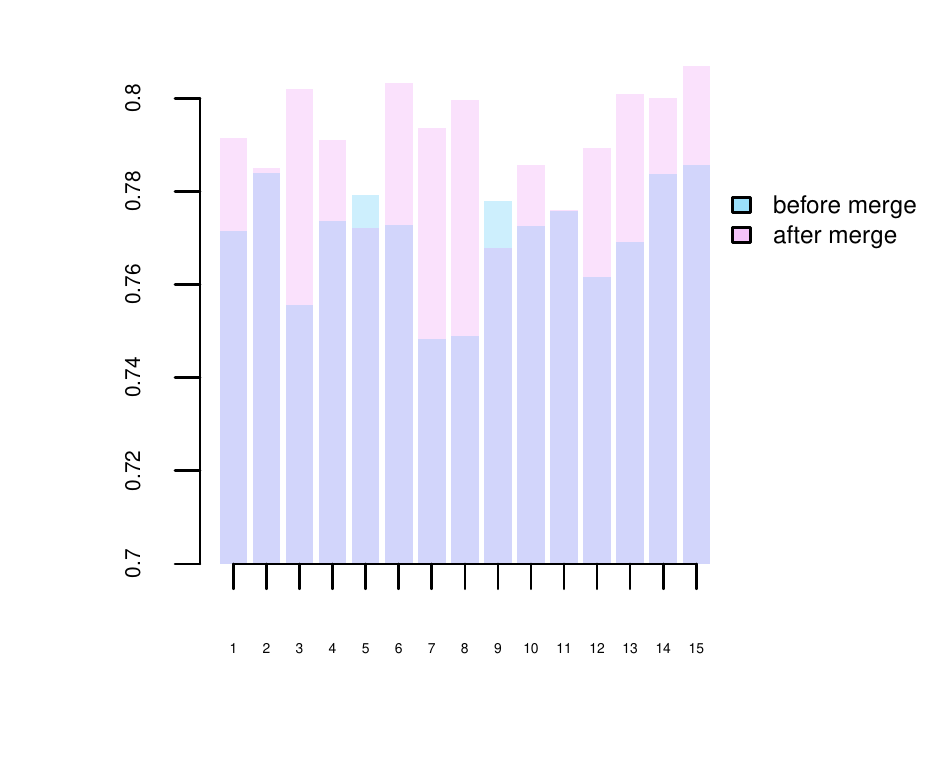}
    \includegraphics[width=0.45\textwidth,height=0.35\textwidth]{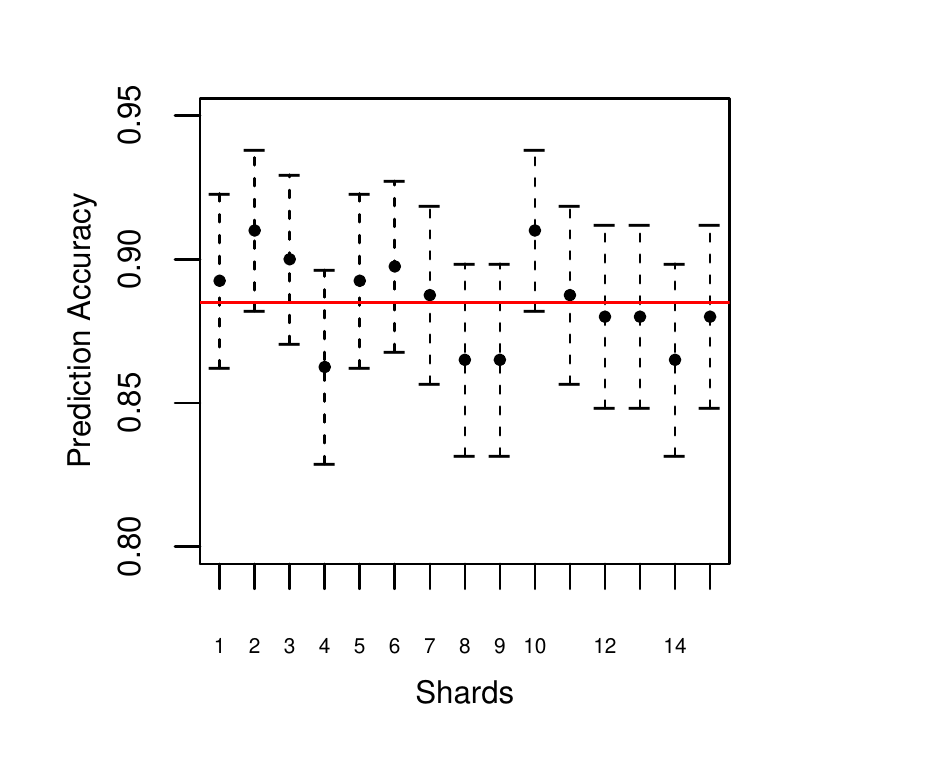}
    \vspace{-5mm}
    \caption{Left: Prediction accuracy pf pairwise comparison in each shard before and after merge. Right: Accuracy  of Level (1-5) prediction of unseen movies  within one level distance in each shard after merge and their $95\%$ confidence intervals. Red line indicates the mean over shards.}
    \label{fig:pair-comp-and-shard-accuracy}
\end{figure}

The resulting average pairwise preference accuracy in each shard is $79.1\%$ with standard deviation $0.012$. \cite{vitelli17mallows} \bch report an accuracy of $79.6\%$ for the pairwise comparison on the same data. \ech  Figure~\ref{fig:pair-comp-and-shard-accuracy} (left panel) shows the before-merge individual shards MCMC pairwise predication accuracy and after-merge pairwise predication accuracy. After-merge prediction accuracy has a higher average and a smaller standard deviation among shards.

For exact predicted rating for unseen movies, on the test set within one star distance, our model has provided $88.5\%$ accuracy. That is saying, we have on average $88.5\%$ probability that our predicted rating is at or within in one star difference of the actual rating.  
Figure~\ref{fig:pair-comp-and-shard-accuracy} (right panel) shows in each shard the prediction accuracy and its $95\%$ Confidence Interval using binomial distribution variance and a normal approximation CI, $(\hat{p}- \mathit{z}_{\alpha} \sqrt{\frac{\hat{p}(1-\hat{p})}{n_s}},\hat{p}+ \mathit{z}_{\alpha} \sqrt{\frac{\hat{p}(1-\hat{p})}{n_s})}$, where $\hat{p}$ denotes the shard accuracy,   $\mathit{z}_{\alpha}$ denotes $1-\frac{\alpha}{2}$ quantile of a standard normal distribution, $\alpha=0.05$.

\begin{figure}
\vspace{-5mm}
    \centering
    \includegraphics[width=\textwidth,height=0.5\textwidth]{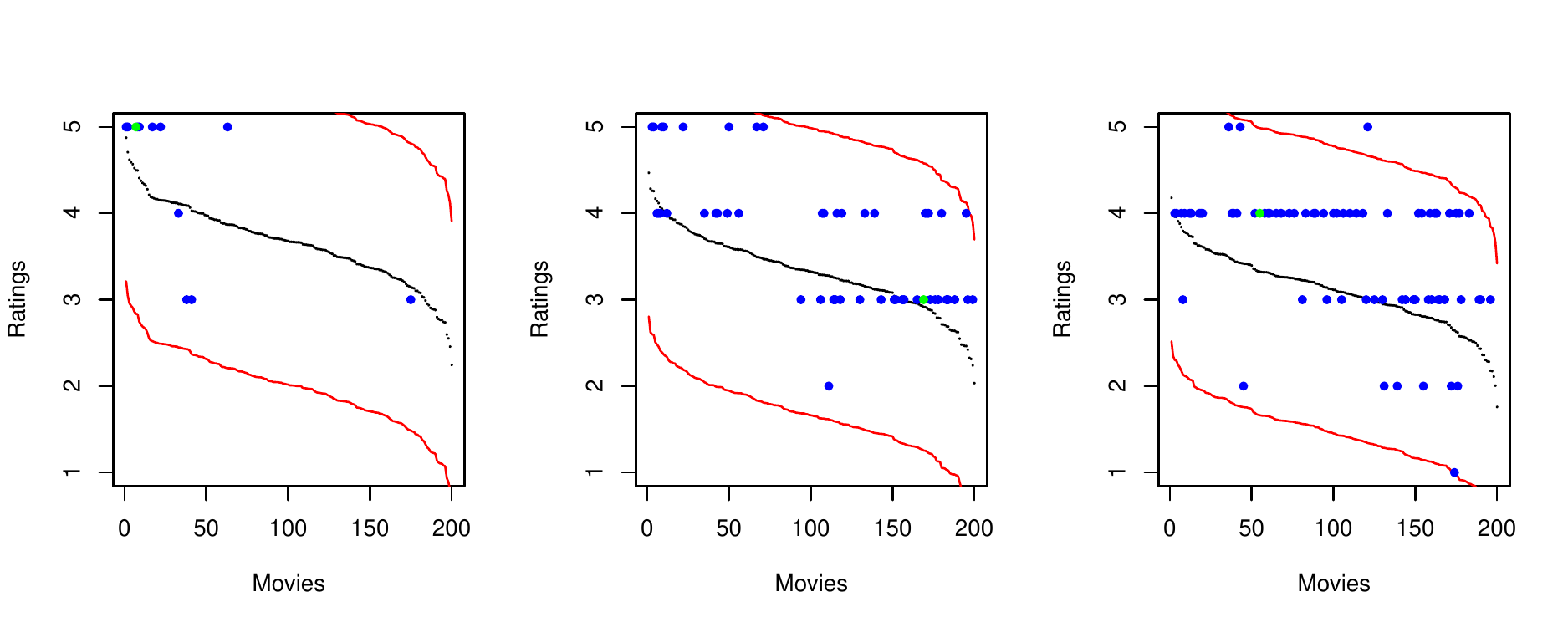}
    \vspace{-5mm}
    \caption{Three randomly selected users' predicted expectation of probit scores for ratings of all movies (seen and unseen), sorted from high to low. The red line indicates the $95\%$ credible intervals. The blue dots are true ratings in training and green dot is the seen test movie rating.}
    \label{fig:three-user-rating}
\end{figure}

\begin{figure}
\vspace{-5mm}
    \centering
    \includegraphics[width=0.5\textwidth]{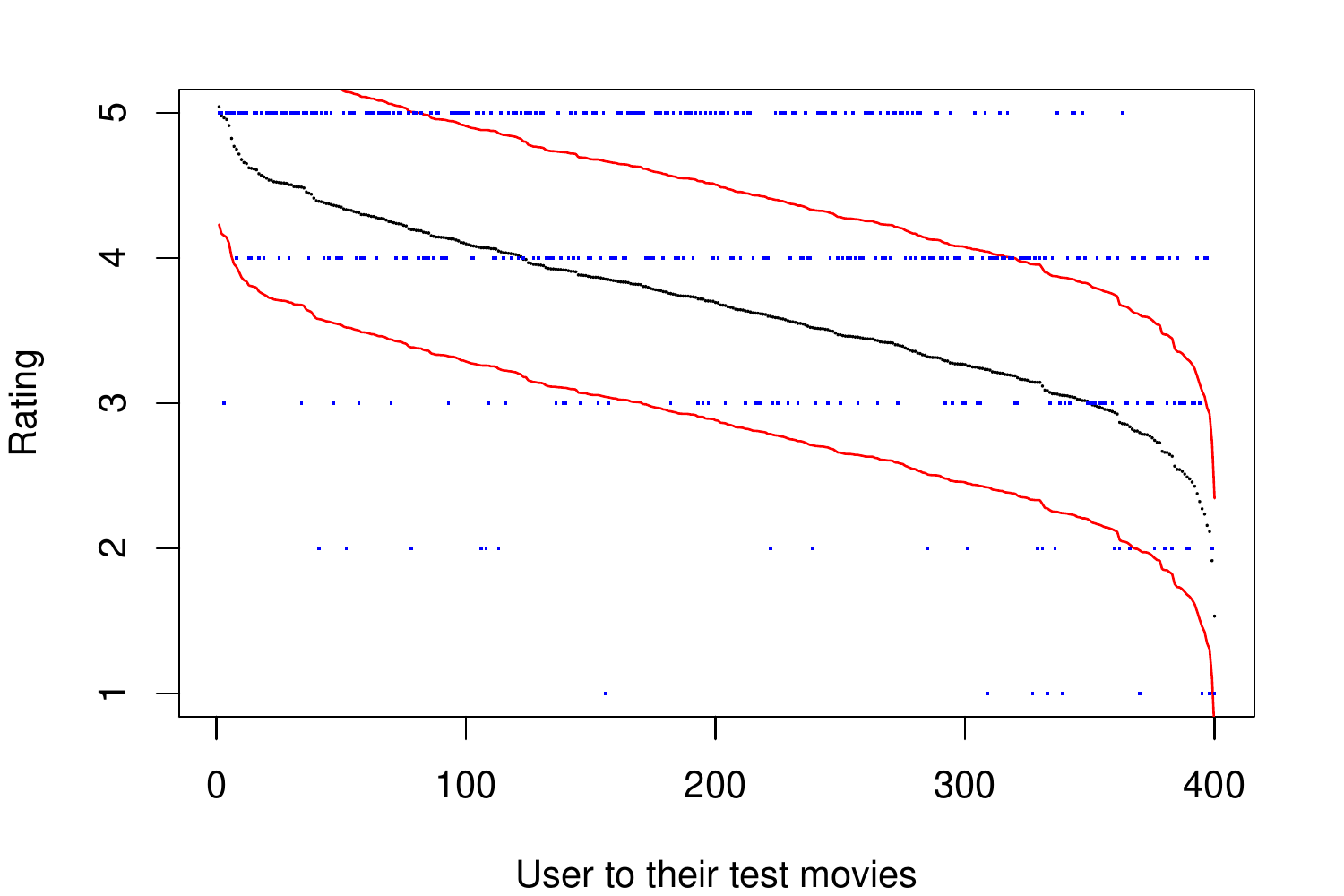}
    \vspace{-3mm}
    \caption{All users' predicted expectation of probit scores for ratings of test movies versus observed ratings of test movies in one randomly picked shard.}
    \label{fig:all-user-testing-in-one-shard}
\end{figure}

We also present individual predicted rating for users. Figure~\ref{fig:three-user-rating} shows three randomly selected user's predicted ratings, $95\%$ credible intervals for all movies (seen and unseen), sorted by predicted ratings, versus true seen ratings of train and test movies. Figure~\ref{fig:all-user-testing-in-one-shard} shows in one randomly picked shard, all users' predicted expectation of probit scores for test movies' ratings versus observed test movies' ratings.

For general rating of all movies, we show the top 10 rated movies across all shards, which is an analysis not available by Mallows Model in \cite{vitelli17mallows}. Table~\ref{tab:top-movies} shows top $10$ movies with highest posterior mean of merged $\rho$.

\begin{table}[]
\scriptsize
\centering
\begin{tabular}{|l|l|l|}
\hline
   & Movie Name                                            & Genre                                                      \\ \hline
1  & The Godfather (1972) & Action/Crime/Drama
                        \\ \hline
2  &  Schindler's List (1993)   &  Drama/War\\ \hline
3  & Star Wars Series                               & Action/Adventure/Drama/Sci-Fi/War                                                \\ \hline
4  & American Beauty (1999)                                 & Comedy/Drama                                           \\ \hline
5  & The Usual Suspects (1995)                          & Crime/Thriller                       \\ \hline
6  & Casablanca (1942)                                  & Drama/Romance/War                                            \\ \hline
7  & Raiders of the Lost Ark (1981)                                         & Action/Adventure                                    \\ \hline
8  &   The Shawshank Redemption (1994)                                   & Drama                                                    \\ \hline
9  & Pulp Fiction (1994)                                    & Crime/Drama                                      \\ \hline
10 & Rear Window (1954)          & Mystery/Thriller                                              \\ \hline
\end{tabular}
\caption{Top $10$ highest rated movies across shards, selected by highest posterior mean of merged $\rho$}
\label{tab:top-movies}
\end{table}

\begin{figure}
\vspace{-5mm}
    \centering
    \includegraphics[width=0.5\textwidth]{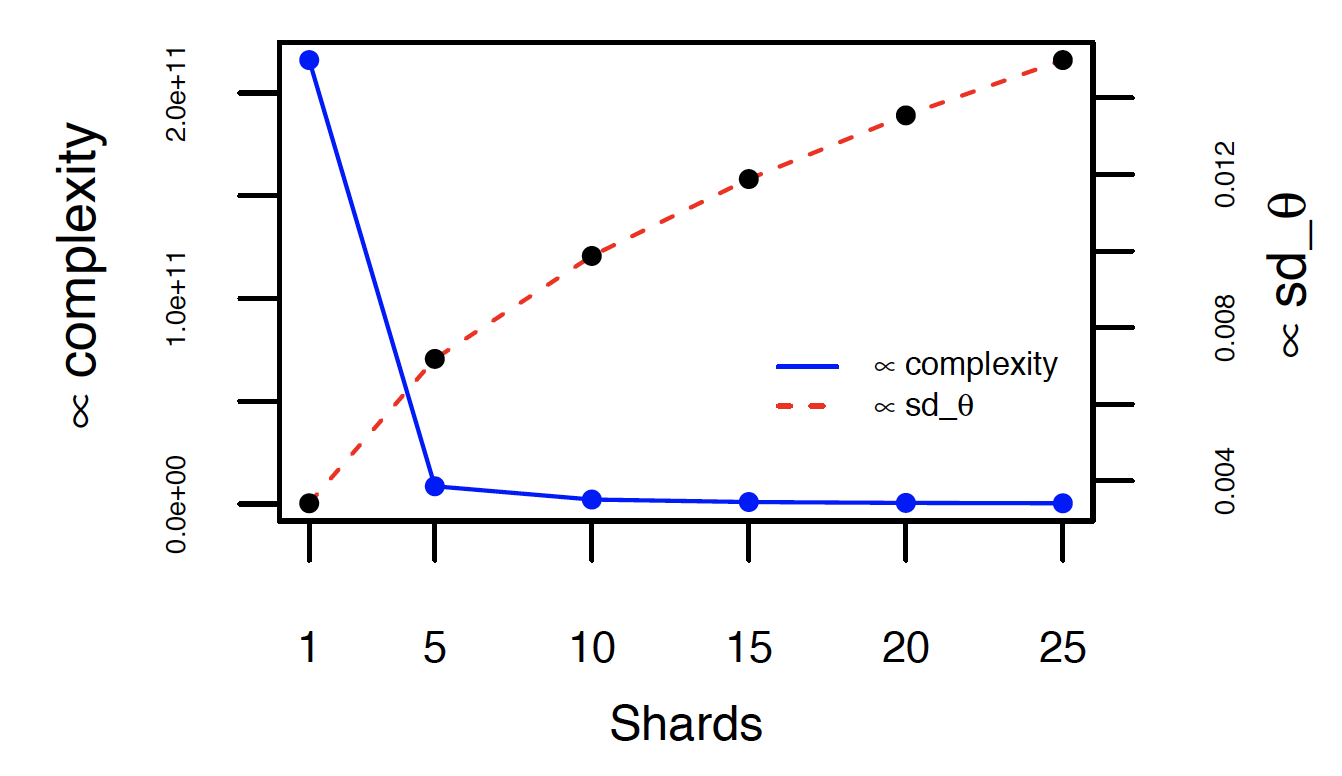}
    \vspace{-3mm}
    \caption{Complexity per iteration and standard error of $\theta$ with the number of shards in consensus monte carlo, $\propto$ denotes proportional up to a constant.}
    \label{fig:comp-sd}
\end{figure}

Finally,  we discuss the computational complexity reduction from our Consensus Monte Carlo method and strategy to choose shard number $S$. \cite{ghahramani2006infinite} has established that an Indian Buffet Process with $N$ data points and linear-Gaussian likelihood model has at least $O(N^3)$ computation complexity per iteration. To be specific, \cite{doshi2009accelerated} pointed out that in each iteration for $N$ number of $D$-dimensional data points and a linear-Gaussian likelihood model, given an $N\times K$ feature assignment matrix in that iteration, the collapsed Gibbs sampler has complexity $O(N^3(K^2+KD))$. In our double feature allocation model with $m\times K$ matrix $A$ and $n\times K$ matrix $B$ ($m>n$) and a probit likelihood model will at least have complexity $O(m^3)$ each iteration. If we split the entire data set into $S$ shards, then for each shard we will have complexity of $O((m/S)^3)$ each iteration, (in total $O(S(m/S)^3)$ for all shards). With $m=6000$ in Movielens data, we see in Figure~\ref{fig:comp-sd} the complexity per iteration decreases and eventually stabilizes with $S$. We also note that more shards would result in fewer users in a shard and fewer ratings observed in each feature $k$ in the shard and thus a larger standard error for $\theta_k$. Consider the normal distribution of $\theta$, heuristically the standard error of $\theta_k$ would be proportional to $\frac{1}{n_{ks}}$ where $n_{ks}$ is the number of ratings observed in feature $k$ in the shard $s$. Following \cite{korwar1973contributions}, the approximate number of features in shard $s$ is $\log(m_sn)$. Suppose $n_{ks}$ is proportional to number of users and movies in the shard and inversely proportional to number of features in shard $s$, i.e., $n_{ks} \propto m_s n(\log(m_sn))^{-1}$, then standard of error of $\theta_k$ is proportional to $((m_s\times n)/\log(m_s\times n))^{-1/2}$. Figure~\ref{fig:comp-sd} shows this trade-off of standard error of estimating $\theta_k$ increasing with number of shards $S$ while the computation cost per iteration decreasing with $S$. Figure~\ref{fig:comp-sd} can serve as a eyeball guideline to pick suitable $S$.

%

\section{Conclusion}
We introduced a novel model-based approach to collaborative filtering,
with the main features being full posterior inference with
interpretable parameters and structure. Based on a generative model, our inference includes a full probabilistic description of any desired summary. For example, the same approach could be used for the performance of learners over problems in a large on-line course. Inference would allow to identify the subsets of similar learners and courses.

Limitations in the current implementation is the very approximate
nature of the reconciliation of the shard-specific posterior
distribution into a reconstructed joint posterior distribution. In
particular, there is no borrowing of information about random subsets
across shards. Such features could be added using, for example, common
anchors in the Consensus Monte Carlo method. An important limitation is the lack of using any covariate information. For example, movies, or generally any items, have known characteristics like actors, length, origin, year etc. Such information could be used to include a rudimentary regression in the subset selection of the generative model. 

Interesting applications arise in many other fields beyond
marketing. For example, users could be HIV patients, items could be
medications and outcomes could any ordinally reported health outcomes,
for example mental health outcomes. Including important baseline
covariates and classes of medications could then allow to recommend
suitable treatment combinations for future patients.

\appendix

\renewcommand{\theequation}{A.\arabic{equation}}
\renewcommand{\thefigure}{A.\arabic{figure}}
\renewcommand{\thetable}{A.\arabic{table}}

\section*{Appendix}
\section{Transition probabilities to update $B$,$\theta$,$\rho$ and $\tau$.}
\begin{itemize}
    \item Update $B|R,A,\mathbf{\theta},\mathbf{\rho},\tau$, \\
    $p(B_{ik}=x|\cdot) \propto p(B_{ik}=x)p(r_{\cdot i}|B_{ik}=x, B_{-i,k},\theta,\rho,\tau, A)$, $x=\{0,1\}$
    
    \item Sample auxiliary variable $\mathbf{Z}|R,A,B,\mathbf{\theta},\mathbf{\rho},\tau$
    $$p(Z_{ui}|\cdot) \sim \mathrm{TruncatedNormal}(b_0+\sum_{k \in K^*}\theta_k+\rho_i,\tau^2) $$
    with lower bound $r_{ui}-1$ ($r_{ui}>0$),  upper bound $r_{ui}$ ($r_{ui}\leq 5$), $K^*$ is the set of $k$ where $A_{uk}=B_{ik}=1$.
    
    \item Update $\mathbf{\tau}|\mathbf{Z},A,B,\mathbf{\theta},\mathbf{\rho}$ from conjuagate Inverse-Gamma families.
    \item Update $\mathbf{\theta}|\mathbf{\rho},\mathbf{Z},\tau,A,B$ and $\mathbf{\rho}|\mathbf{\theta},\mathbf{Z},\tau,A,B$ from conjuagate Normal families. 
     
\end{itemize}
\clearpage
\bibliography{references}
\end{document}